\documentclass[a4paper,10pt]{article}
\usepackage{jcappub}
\usepackage[T1]{fontenc} 
\input psfig.sty

\title{\boldmath Constraining the local variance of $H_0$ from directional analyses}

\author{C. A. P. Bengaly, Jr.}
\affiliation[a]{Observat\'orio Nacional, 20921-400, Rio de Janeiro - RJ, Brasil}
\emailAdd{carlosap@on.br}

\abstract{We evaluate the local variance of the Hubble Constant $H_0$ with low-z Type Ia Supernovae (SNe). Our analyses are performed using a hemispherical comparison method in order to test whether taking the bulk flow motion into account can reconcile the measurement of the Hubble Constant $H_0$ from standard candles ($H_0 = 73.8 \pm 2.4 \; \mathrm{km \; s}^{-1}\; \mathrm{Mpc}^{-1}$) with that of the Planck's Cosmic Microwave Background data ($H_0 = 67.8 \pm 0.9 \; \mathrm{km \; s}^{-1} \mathrm{Mpc}^{-1}$). We obtain that $H_0$ ranges from $68.9 \pm 0.5 \; \mathrm{km \; s}^{-1} \mathrm{Mpc}^{-1}$ to $71.2 \pm 0.7 \; \mathrm{km \; s}^{-1} \mathrm{Mpc}^{-1}$ through the celestial sphere ($1\sigma$ uncertainty), implying a Hubble Constant maximal variance of $\delta H_0 = (2.30 \pm 0.86) \; \mathrm{km \; s}^{-1} \mathrm{Mpc}^{-1}$ towards the $(l,b) = (315^{\circ},27^{\circ})$ direction. Interestingly, this result agrees with the bulk flow direction estimates found in the literature, as well as previous evaluations of the $H_0$ variance due to the presence of nearby inhomogeneities. We assess the statistical significance of this result with different prescriptions of Monte Carlo simulations, obtaining moderate statistical significance, i.e., $68.7$\% confidence level (CL) for such variance. Furthermore, we test the hypothesis of a higher $H_0$ value in the presence of a bulk flow velocity dipole, finding some evidence for this result which, however, cannot be claimed to be significant due to the current large uncertainty in the SNe distance modulus. Then, we conclude that the tension between different $H_0$ determinations can plausibly be caused to the bulk flow motion of the local Universe, even though the current incompleteness of the SNe data set, both in terms of celestial coverage and distance uncertainties, does not allow a high statistical significance for these results or a definitive conclusion about this issue.}

\begin{document}
\maketitle
\flushbottom

%%%%%%%%%%%%%%%%%%%%%%%%%%%%%%%%%%%%%%%%%%%%%%%%
\section{Introduction}    \label{intro}
%%%%%%%%%%%%%%%%%%%%%%%%%%%%%%%%%%%%%%%%%%%%%%%%

For over a decade, the $\Lambda$CDM scenario has been consolidated as the concordance model of cosmology. Since the pioneering works of~\cite{riess1, perl}, who showed that the cosmological expansion is incompatible with a null Cosmological Constant $\Lambda$ value at more than $2\sigma$ confidence level (CL), this paradigm has been exhaustively tested with larger and more precise SNe catalogues~\cite{U2.1, JLA}, Cosmic Microwave Background (CMB) temperature fluctuations~\cite{planck13, planck15}, large-scale structure information~\cite{boss}, as well as ages of old cosmic objects~\cite{simon, stern, moresco}, being the most successful proposal to describe the observations until the present moment. A fundamental cosmological quantity of the concordance model is the Hubble Constant, $H_0$, which gives information about the current expansion rate of the Universe. There are attempts to evaluate this parameter since the early twentieth century, and in the recent years it turned out to be possible to determine it within an uncertainty of a few per cent, as achieved by~\cite{planck15} and~\cite{riess2}.

However, there is a recent debate in the literature on the conflict between independent $H_0$ measurements. Whereas~\cite{riess2} obtained $73.8 \pm 2.4 \; \mathrm{km \; s}^{-1}\; \mathrm{Mpc}^{-1}$ from the magnitude of standard candles, the Planck team~\cite{planck13} estimated a lower value for it, $67.3 \pm 1.2\; \mathrm{km \; s}^{-1} \mathrm{Mpc}^{-1}$, whose 2015 update provided $67.8 \pm 0.9\; \mathrm{km \; s}^{-1} \mathrm{Mpc}^{-1}$~\cite{planck15}, thus corresponding to a $\sim 2.5\sigma$ tension. Since then, many attempts were carried out to reconcile this discrepancy. For instance,~\cite{efstathiou} re-analysed the Cepheid data of~\cite{riess2} and obtained $H_0 = 72.5 \pm 2.5 \; \mathrm{km \; s}^{-1} \mathrm{Mpc}^{-1}$, reducing the tension to $\sim 1.9\sigma$. Model-independent tests were also developed to estimate $H_0$, such as in~\cite{busti}, where a non-parametric reconstruction of the $H(z)$ data indicates an even lower Hubble Constant value, $H_0 = 64.9 \pm 4.2\; \mathrm{km \; s}^{-1} \mathrm{Mpc}^{-1}$, than the Planck's determination although compatible within $1\sigma$ CL with it. In addition,~\cite{wu} also obtained a $H_0$ measurement compatible with the CMB one from the cosmic duality distance relation test. 

One possible explanation for such tension consists on the assumption that a different expansion rate might occur because of the gravitational instability related to the inhomogeneities of the local cosmic web. This was analysed by~\cite{marra, dayan}, where they calculated the fluctuation of the $H_0$ value due to the cosmic variance and found out that it presents the same magnitude as the Hubble Constant observational uncertainty though does not completely solve this tension. N-body simulations of the low-z cosmic web were also performed by~\cite{wojtak, odderskov1, odderskov2, odderskov3}, but the variance obtained in the Hubble Constant is too small to explain the conflict. Finally, there are also earlier works concerning backreaction effects of nearby structures on the inference of global cosmological parameters, such as~\cite{li, wiegand}, but again the $H_0$ fluctuation estimated is not large enough to reconcile these discrepant determinations.   

Distance probes in the low redshift range, such as galaxies via Tully-Fisher relation and SNe distance moduli, have been adopted for a long time to study the possibility of bulk flow motion due to the local inhomogeneities, which is manifested as a velocity dipole correction in the distance of these objects. Many works were dedicated to its characterisation, as in~\cite{colin, dai, ma1, turn, rath, feindt, ma2, app1, app2, math, huterer}, who used SNe to probe this possible bulk flow motion, in addition to those adopting low-z galaxy catalogues~\cite{nuss, branch, wilt, hong, watk2, mckay, bolejko}, a combination of both cosmic artifacts~\cite{watk1, feld}, as well as galaxy clusters~\cite{kash1, kash2, mak, osb, atrio}. These analyses lead to quite similar results for the bulk flow direction, although the amplitude is still a matter of some debate\footnote{The analyses with galaxy clusters point out a very large flow (about $\sim 1000 \; \mathrm{km \; s}^{-1}$) via kinematic Sunyaev-Zel'dovich effect, and also that it does not converge even in very large scales ($\simeq 800$ Mpc).}. Moreover,~\cite{schwarz, kalus, chang, bengaly} analysed the $H_0$ spatial variation with low-z SNe data in order to test the global cosmological isotropy, finding a potential correlation between its directional asymmetry and the bulk flow direction which raises the hypothesis that the low and high-z $H_0$ discrepancy could be related to this phenomenon. This idea was also explored by~\cite{tsagas1, tsagas2, tsagas3}, who discussed, from a more theoretical framework, that we could be experiencing an accelerating Universe due to this large scale flow, and that such an effect would be revealed in the form of a dipole in the cosmological expansion. However, observational constraints by~\cite{bengaly, antoniou, cai, jimenez, javanmardi} are still ambiguous on the significance of this anisotropic signal.  

Therefore, it is of great importance to probe the variance of $H_0$ with respect to the local cosmic web, which we do so by constraining its directional asymmetry through the entire celestial sphere. For this purpose, nearby SNe ($z \leq 0.1$) from the Union2.1 catalogue~\cite{U2.1} encompassed in opposite hemispheres~\cite{bernui} are adopted. We investigate whether this asymmetry is consistent with the bulk flow direction reported in the literature, and whether the difference between $H_0$ estimated in antipodal patches of the sky can lead to a variance that can possibly explain the tension discussed above. Moreover, we stress that our goal is to perform an evaluation of the Hubble Constant variance essentially in terms of observational information and statistical significance assessment, using the results present in the literature for the bulk flow motion for a consistency check. Hence, the structure of the paper is the following: Section 2 describes the methods developed to perform our anisotropic studies and its statistical significance. Section 3 is dedicated to analyses of our results, including the statistical significance tests and, finally, the conclusions and discussions of this work are presented in section 4.

%%%%%%%%%%%%%%%%%%%%%%%%%%%%%%%%%%%%%%%%%%%%%%%%
\section{The methodology}   \label{method}
%%%%%%%%%%%%%%%%%%%%%%%%%%%%%%%%%%%%%%%%%%%%%%%%

\subsection{Mapping the Hubble Constant} 

In our analyses, the local variance of $H_0$ is estimated by constructing a map of this quantity through the celestial sphere. We employ the opposite hemisphere method to accomplish this, following the same approach as in~\cite{bengaly} (see also~\cite{bernui}). In this case, the centre of each hemisphere is defined by the HEALPix pixelisation scheme~\cite{gorski}, so that we estimate $H_0$ from the SNe comprised in these hemispheres following a $\chi^2$ fitting procedure according to 

\begin{equation}
\label{eq:chi2}
\chi^2 = \sum_i\left(\frac{\mu_i-\mu_{\mathrm{th}}(z_i,\mathbf{p})}{\sigma_{\mu_i}}\right)^2 \;,
\end{equation}

\noindent where the set $(z_i, \mu_i, \sigma_{\mu_i})$ is the observational information of the SNe data, i.e., redshift, distance modulus and associated uncertainty of the $i$-th object, respectively, and $\mu_{\mathrm{th}}(z,\mathbf{p})$ is the distance modulus given by

\begin{equation}
\label{eq:mu_th}
\mu_{\mathrm{th}}(z_i,\mathbf{p}) = m - M = 5\log_{10}{ \left[ \frac{D_L(z_i,\mathbf{p})}{Mpc} \right] } + 25 \;,
\end{equation}

\noindent where $D_L(z_i,\mathbf{p})$ is the luminosity distance, whose arguments are the redshift $z$ of the $i$-th SN and the set of cosmological parameters $\mathbf{p}$ which describe the underlying cosmological model. As our analyses are restricted to a low-z regime, $D_L(z,\mathbf{p})$ is well described by the following cosmographic expansion

\begin{equation}
\label{eq:DL}
D_L(z,\mathbf{p}) = \frac{c}{H_0} \left[ z + \frac{ (1 - q_0)z^2 }{2} \right]\;,
\end{equation}

\noindent whose parameter $q_0$ represents the deceleration parameter. Throughout this paper, we set $q_0 = -0.5845$, which is the exact value for this parameter when $\Omega_m = 0.277$ is assumed, corresponding to the Union2.1 best fit for a flat $\Lambda$CDM cosmological model. We must stress that assuming a fixed value of $q_0$ does not provide any loss of generality in our analyses due to the low sensitivity of $D_L$ on the $q_0$ parameter in the chosen redshift range, therefore different values for $q_0$ do not significantly affect our results. As the data consist of the SNe encompassed in each hemisphere, we can assign the best-fit $H_0$ to the pixel corresponding to the centre of these hemispheres, thus constructing a map of directional dependence of the Hubble Constant which we call {\it Hubble-map} throughout this paper\footnote{Unless stated otherwise, we choose $N_{\mathrm side} = 8$ ($7.33^{\circ}$ angular resolution) pixelisation scheme for our analyses, so that we have 768 available hemispheres for our analyses.} . 

\begin{figure*}%[!t]
\includegraphics[width = 6.0cm, height = 7.5cm, angle = +90]
{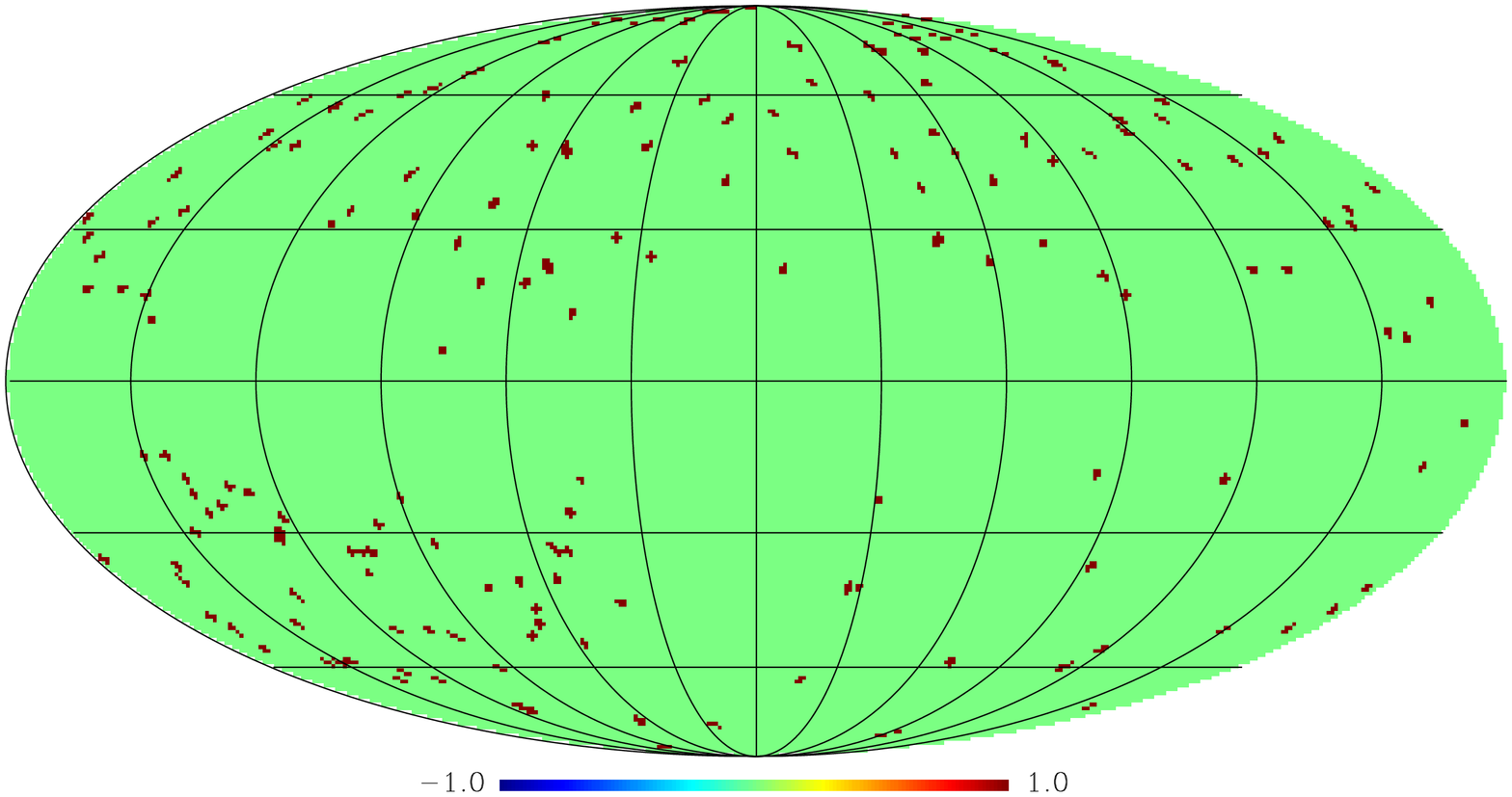}
\hspace{0.3cm}
\includegraphics[width = 6.0cm, height = 7.5cm, angle = +90]
{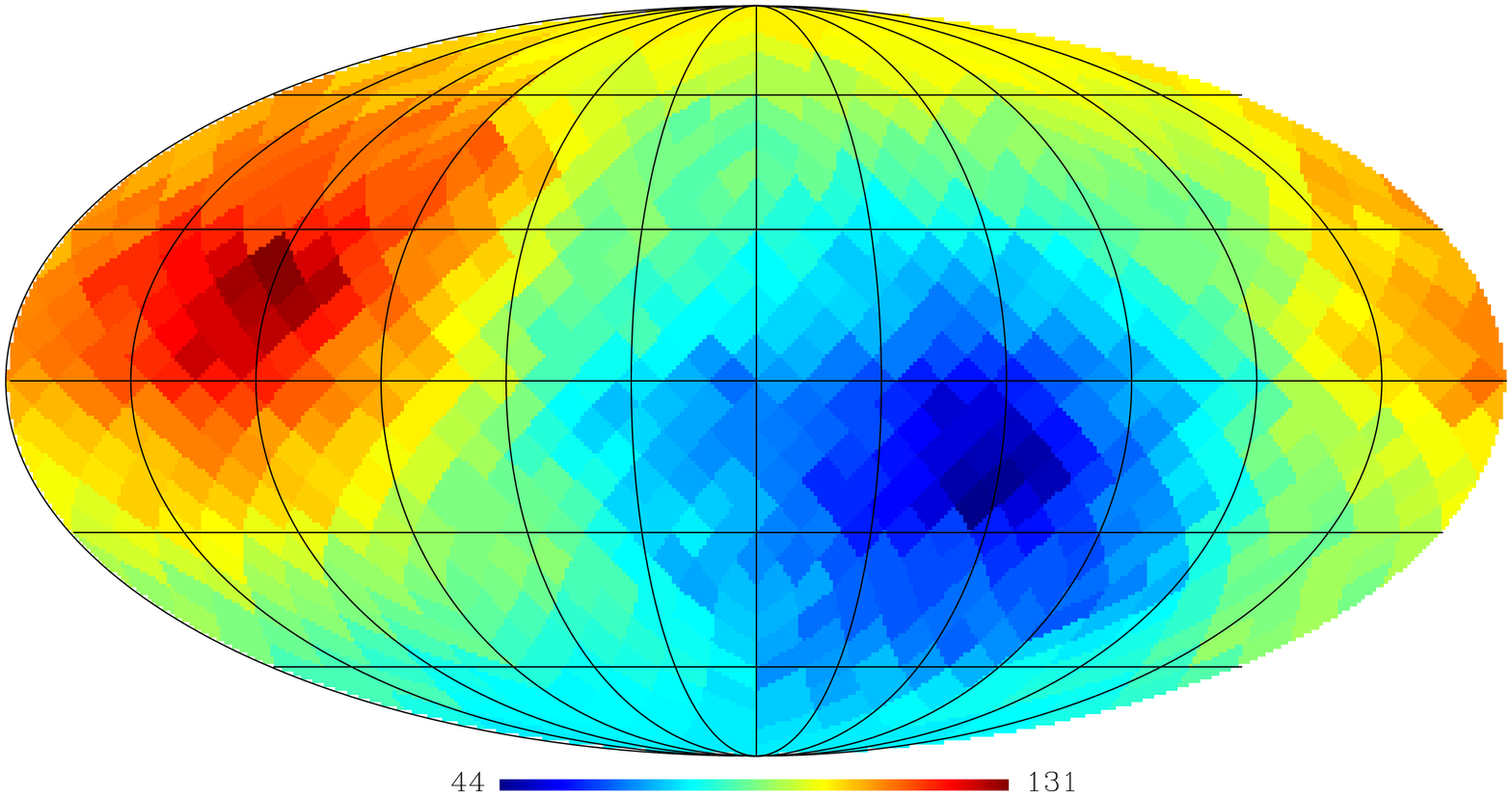}
\caption{ {\it Left panel:} The angular distribution of the Union2.1 SNe events in the $z \leq 0.10$ range. {\it Right panel:} the number of these objects encompassed in each of the hemispheres available for our analyses. It ranges from 44 to 131, where 175 is the total number of objects present in this sample.} 
\label{fig:SNe_maps}
\end{figure*}

\subsection{Statistical significance tests} 

In order to assess the statistical significance of the Hubble-map analysis, we produce 1000 Monte Carlo (MC) realisations with four different prescriptions, which consist in: 

\begin{itemize}

\item {\it MC-shuffle}: The angular coordinates of each SN is maintained, though their $(z, \mu, \sigma_{\mu})$ is shuffled.

\item {\it MC-iso}: The SNe data are isotropically redistributed on the celestial sphere.

\item {\it MC-gauss-aniso}: The original positions of each SNe in the sky is maintained, while their distance moduli is drawn from a Gaussian distribution centred at $\mu_{fid}$, i.e., a fiducial value computed according to $H_0 = 70 \; \mathrm{km \; s}^{-1} \mathrm{Mpc}^{-1}$ and $q_0 = -0.5845$ at each redshift, whose standard deviation is given by the $\sigma_{\mu}$ of the corresponding SN.

\item {\it MC-gauss-iso}: Same as the third test, but redistributing the objects isotropically as in the second prescription.

\end{itemize}

The goal of the first two tests is to compute the statistical significance of the Hubble-map analysis according to the distribution pattern in the sky of the Union2.1 data points, while the third and fourth MC ensembles estimate how the uncertainty of the distance measurements affects the statistical significance of the results. In addition, we also verify whether the presence of a bulk flow velocity actually leads to a bias in the $H_0$ measurement. This is assessed via MC realisations similarly to the MC-gaussian-aniso test discussed before, however, this time we carry out a global fit for the Hubble constant value instead of estimating its value through different patches of the sky, and we take the bulk flow motion into account. This effect is computed following~\cite{bonvin}, who modeled it as a dipole correction term in eq.~\ref{eq:DL} according to

\begin{equation}
\label{eq:DL_dip}
D^{(dip)}_L(z,\mathbf{p},\theta) = \frac{v_{bf}(1+z)^2}{H(z)}\cos{\theta}\;,
\end{equation}

\noindent where $H(z) = H_0[1 + (1 + q_0)z]$, $v_{bf}$ is the bulk flow velocity, and $\theta$ represents the angle between the position of the corresponding SN and the bulk flow direction. 

\begin{figure*}%[!t]
\includegraphics[width = 6.0cm, height = 7.5cm, angle = +90]
{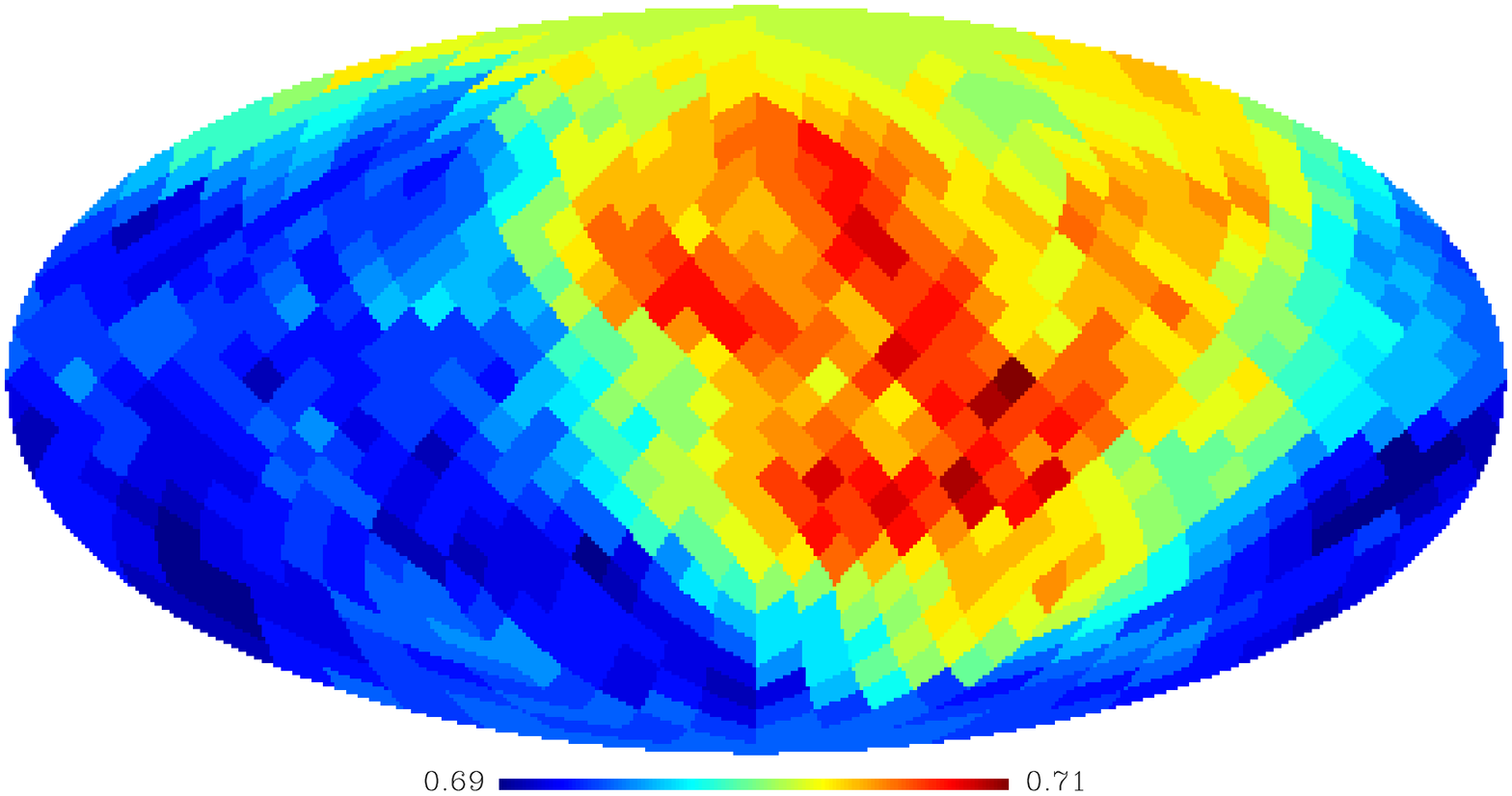}
\hspace{0.3cm}
\includegraphics[width = 6.0cm, height = 7.5cm, angle = +90]
{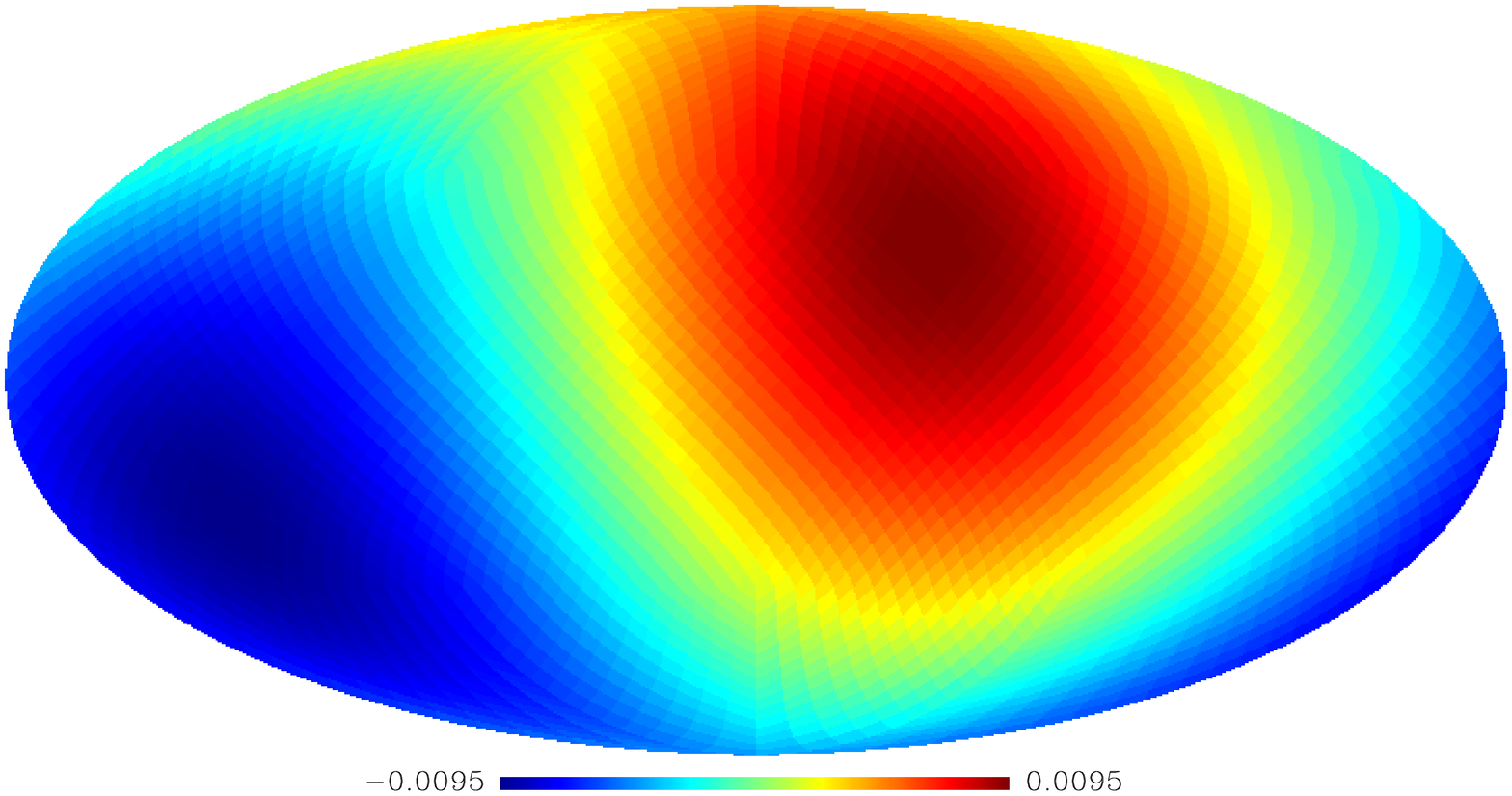}
\caption{ {\it Left panel}: The result of the {\it Hubble-map} analysis given in terms of $H_0/100$. The lowest and highest values obtained are $H_0 = 68.9 \pm 0.5 \; \mathrm{km \; s}^{-1} \mathrm{Mpc}^{-1}$ and $71.2 \pm 0.7 \; \mathrm{km \; s}^{-1} \mathrm{Mpc}^{-1}$ ($1\sigma$ uncertainty), respectively, yielding $\delta H_0 = (2.30 \pm 0.86) \; \mathrm{km \; s}^{-1} \mathrm{Mpc}^{-1}$. {\it Right panel}: The dipole-only contribution of the Hubble-map, given in arbitrary units and $N_{side} = 16$, whose axis of maximal $H_0$ variance points toward $(l,b) = (315.00^{\circ},27.28^{\circ})$.} 
\label{fig:hubble_maps}
\end{figure*}

%%%%%%%%%%%%%%%%%%%%%%%%%%%%%%%%%%%%%%%%%%%%%%%%
\section{Results}   \label{results}
%%%%%%%%%%%%%%%%%%%%%%%%%%%%%%%%%%%%%%%%%%%%%%%%

\subsection{The Hubble-map}

\begin{table}[!t]
\begin{center}
\begin{tabular}{ccccc}
\hline
\hline
Redshift range & $\delta H_0 \; (\mathrm{km \; s}^{-1} \mathrm{Mpc}^{-1})$ & $(l,b)$ & Ref.\\
\hline
\hline
$0.01 \leq z \leq 0.10$ & $1.79$ & - & \cite{marra} \\
$ z \leq 0.10$ & $1.6 - 2.4$ & - & \cite{dayan} \\
$z \leq 0.20$ & $3.0$ & $(302^{\circ},-18^{\circ})$ & \cite{kalus} \\
$z \leq 0.20$ & $4.6$ & $(326^{\circ},12^{\circ})$ & \cite{bengaly} \\
\hline
\hline
$z \leq 0.10$ & $2.30 \pm 0.86$ & $(315^{\circ},27^{\circ})$ & This work \\
\hline
\hline
Redshift range & $v_{bf} \; (\mathrm{km \; s}^{-1}$) & $(l,b)$ & Ref.\\
\hline
\hline
$z < 0.06$ & $260 \pm 130$ & $(282^{\circ} \pm 34^{\circ}, 22^{\circ} \pm 20^{\circ})$ & \cite{colin} \\
$z < 0.05$ & $188 \pm 120$ & $(298^{\circ} \pm 40^{\circ}, 8^{\circ} \pm 40^{\circ})$ & \cite{dai} \\
$z < 0.20$ & $260$ & $(295^{\circ}, 5^{\circ})$ & \cite{rath} \\
$0.015 < z < 0.035$ & $292 \pm 96$ & $(290^{\circ} \pm 22^{\circ}, 15^{\circ} \pm 18^{\circ})$ & \cite{feindt} \\
$z < 0.05$ & $270 \pm 50$ & $(295^{\circ} \pm 30^{\circ}, 10^{\circ} \pm 15^{\circ})$ & \cite{math} \\
$0.015 < z < 0.045$ & - & $(276^{\circ} \pm 29^{\circ}, 20^{\circ} \pm 14^{\circ})$ & \cite{app2} \\
\hline
\hline
\end{tabular}
\end{center}
\caption{A summary of the $H_0$ variance estimates, as well as the estimation of bulk flow velocity and direction, reported in the literature compared to our own.} 
\label{tab:tab_deltah_vbf} 
\end{table}

The left panel of figure~\ref{fig:SNe_maps} shows the angular distribution of SN events at $z \leq 0.1$ for the Union2.1~\cite{U2.1} compilation, while the number of objects encompassed in each hemisphere is exhibited on the right panel of the same figure. This number ranges from 44 to 131 through the celestial sphere, which evidences the non-uniformity of the SNe angular distribution pattern. Nevertheless, the Union2.1 angular distribution is similar to that of the Constitution SNe compilation~\cite{hicken}, i.e., the data that was adopted by~\cite{riess2} to perform their $H_0$ estimation\footnote{This can be seen comparing figure 8 of~\cite{kalus}, who used the Constitution data set for their analyses, with the right panel of Fig.~\ref{fig:SNe_maps} of this paper.}.

The result of the Hubble-map analysis is presented in the left panel of figure~\ref{fig:hubble_maps}. We obtain a lowest $H_0$ estimate of $68.9 \pm 0.5 \; \mathrm{km \; s}^{-1} \mathrm{Mpc}^{-1}$, and a highest estimate of $71.2 \pm 0.7 \; \mathrm{km \; s}^{-1} \mathrm{Mpc}^{-1}$, hence the amplitude of its variance is $\delta H_0 = (2.30 \pm 0.86) \; \mathrm{km \; s}^{-1} \mathrm{Mpc}^{-1}$. We note that these values agree with both $H_0$ measurements of~\cite{planck15} and~\cite{riess2}, respectively, assuming $1\sigma$ CL. Besides, this $\delta H_0$ is in good agreement with most of the estimates reported in the literature, as those listed in the table~\ref{tab:tab_deltah_vbf}. For instance,~\cite{kalus} obtained a similar result, even though they probed $\delta H_0$ to a higher redshift than we did. On the other hand, the analysis performed in~\cite{bengaly} lead to the highest $\delta H_0$ of all the current estimations since the authors marginalised $q_0$ instead of setting this parameter to a specific value. Interestingly, our result is also in concordance with the~\cite{marra} estimate of $\delta H_0 = 1.79 \; \mathrm{km \; s}^{-1} \mathrm{Mpc}^{-1}$ due to the cosmic variance related to nearby inhomogeneities, as well as~\cite{dayan}, who obtained that $1.6 \leq \delta H_0 \leq 2.4 \; \mathrm{km \; s}^{-1} \mathrm{Mpc}^{-1}$ from a fully relativistic cosmic variance estimation, both in the same redshift range as our analyses. We also compute the direction of the Hubble Constant maximal variance from the dipole of the Hubble-map analysis, which is shown in the right panel of figure~\ref{fig:hubble_maps}. We find that the maximal dipolar signal points toward the $(l,b) = (315.00^{\circ},27.28^{\circ})$ direction, hence it is in good agreement with the reports of the bulk flow motion direction presented in the table~\ref{tab:tab_deltah_vbf}.

\subsection{Statistical significance tests} 

\begin{figure*}[!t]
\includegraphics[scale = 0.45]
{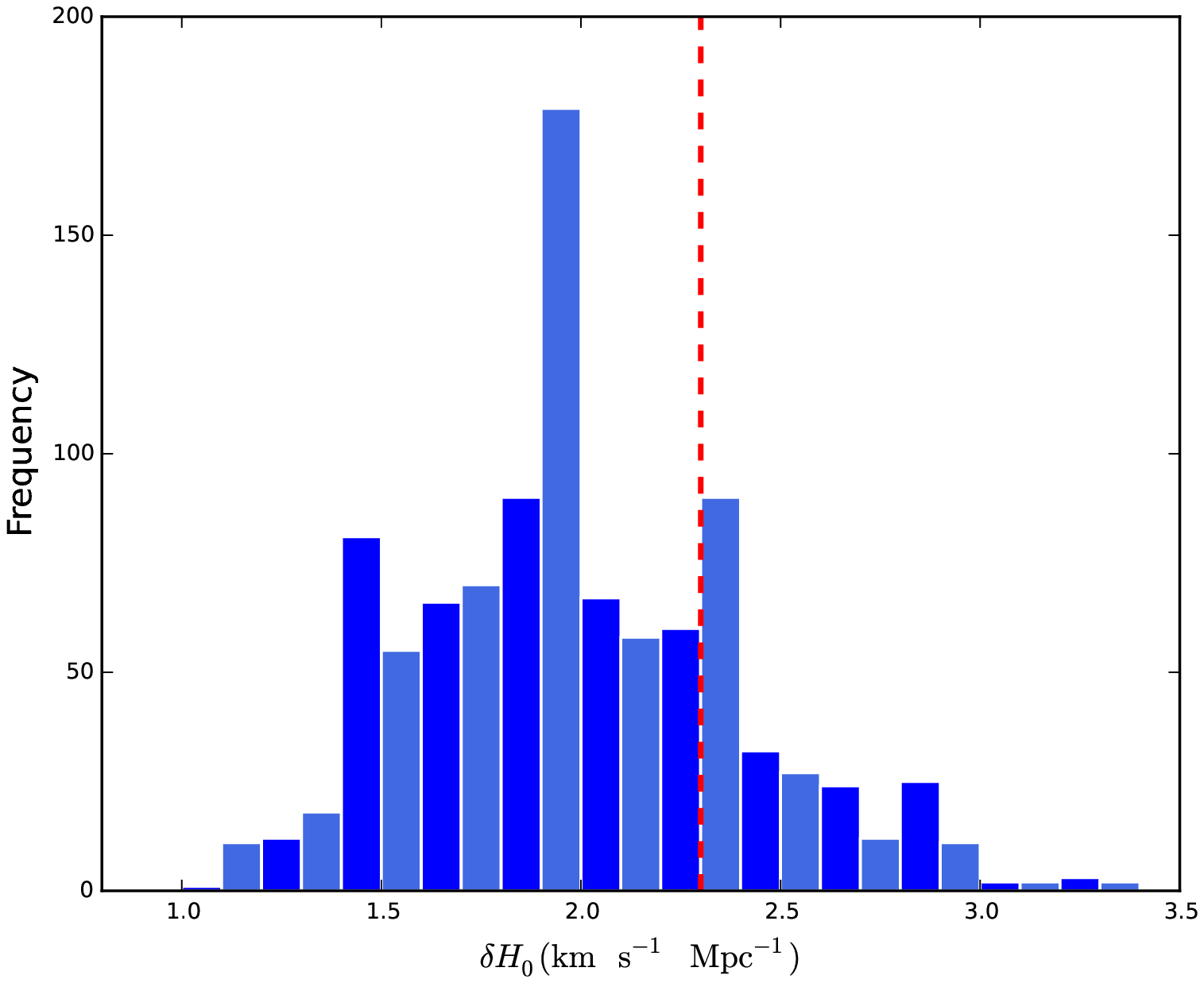}
\hspace{0.3cm}
\includegraphics[scale = 0.45]
{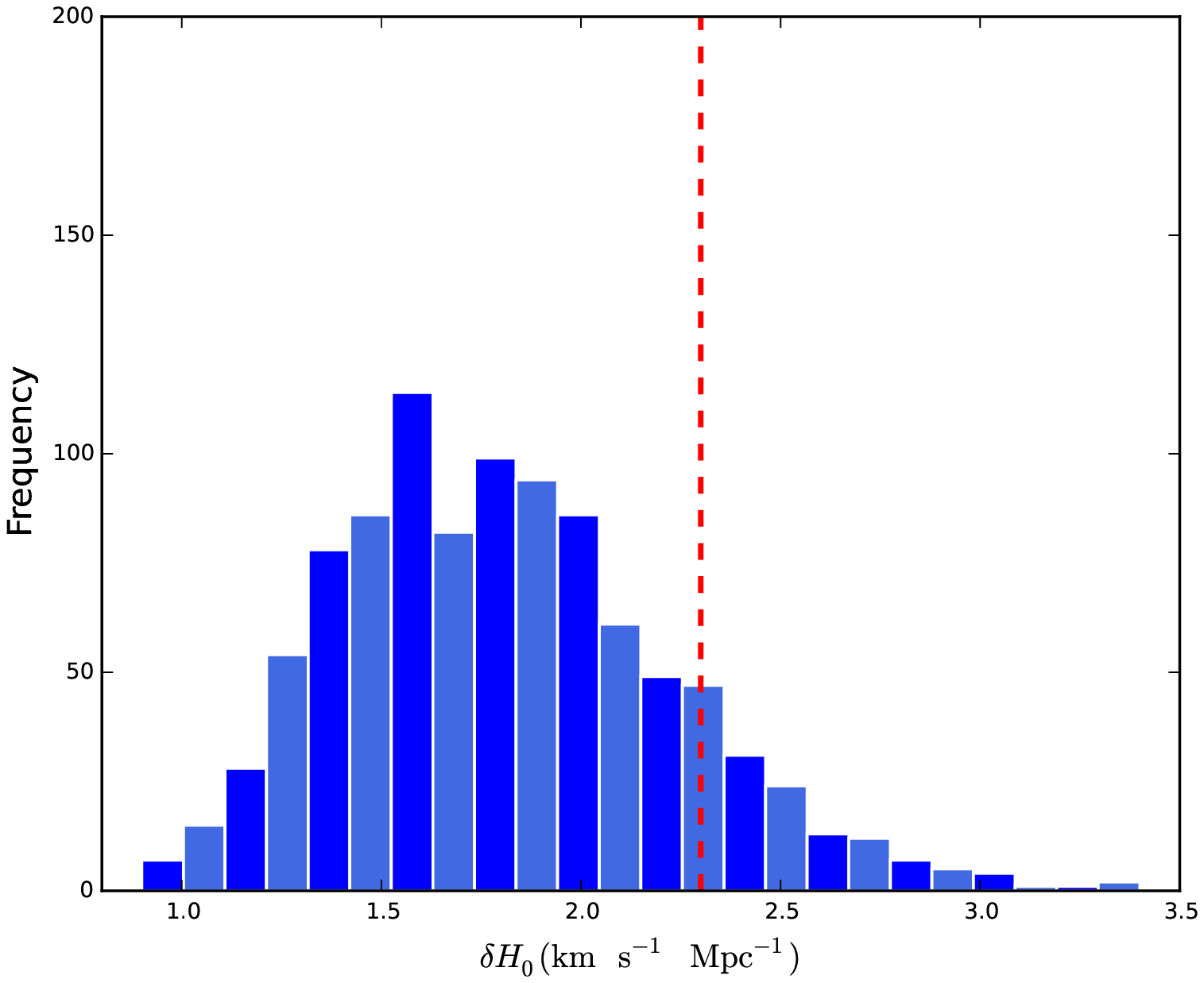}
\caption{{\it Left panel}: The histogram of the $\delta H_0$ provided by 1000 {\it MC-shuffle} simulations. {\it Right panel}: the same histogram, but for the {\it MC-iso} simulations. We find that the variance $\delta H_0 = 2.3 \; \mathrm{km \; s}^{-1} \mathrm{Mpc}^{-1}$, highlighted in the red dashed vertical line, or larger can be reproduced by $20.4$\% and $14.7$\% of these runs, respectively.} 
\label{fig:histograms_deltah_01}
\end{figure*}

\begin{figure*}[!t]
\includegraphics[scale = 0.45]
{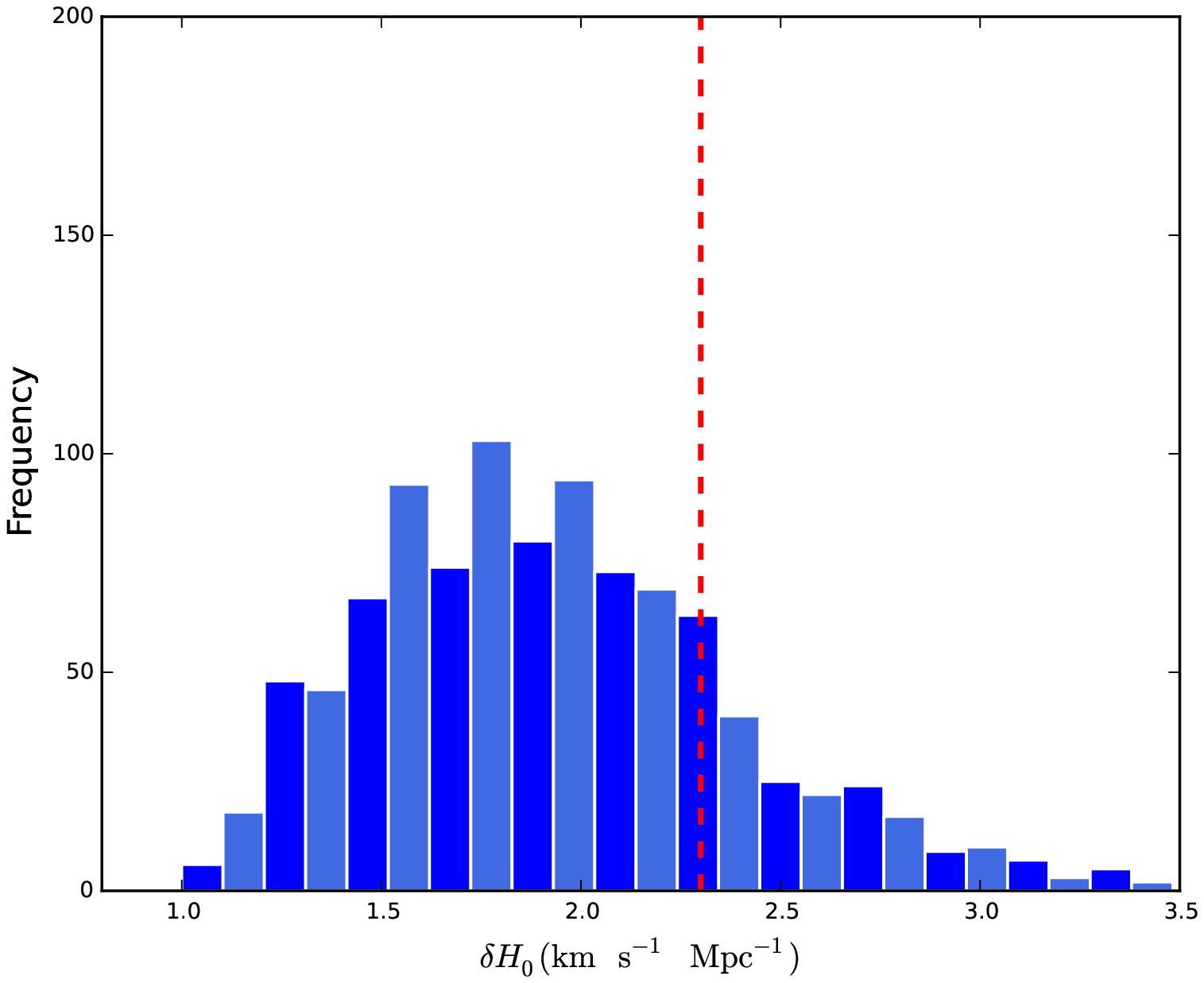}
\hspace{0.3cm}
\includegraphics[scale = 0.45]
{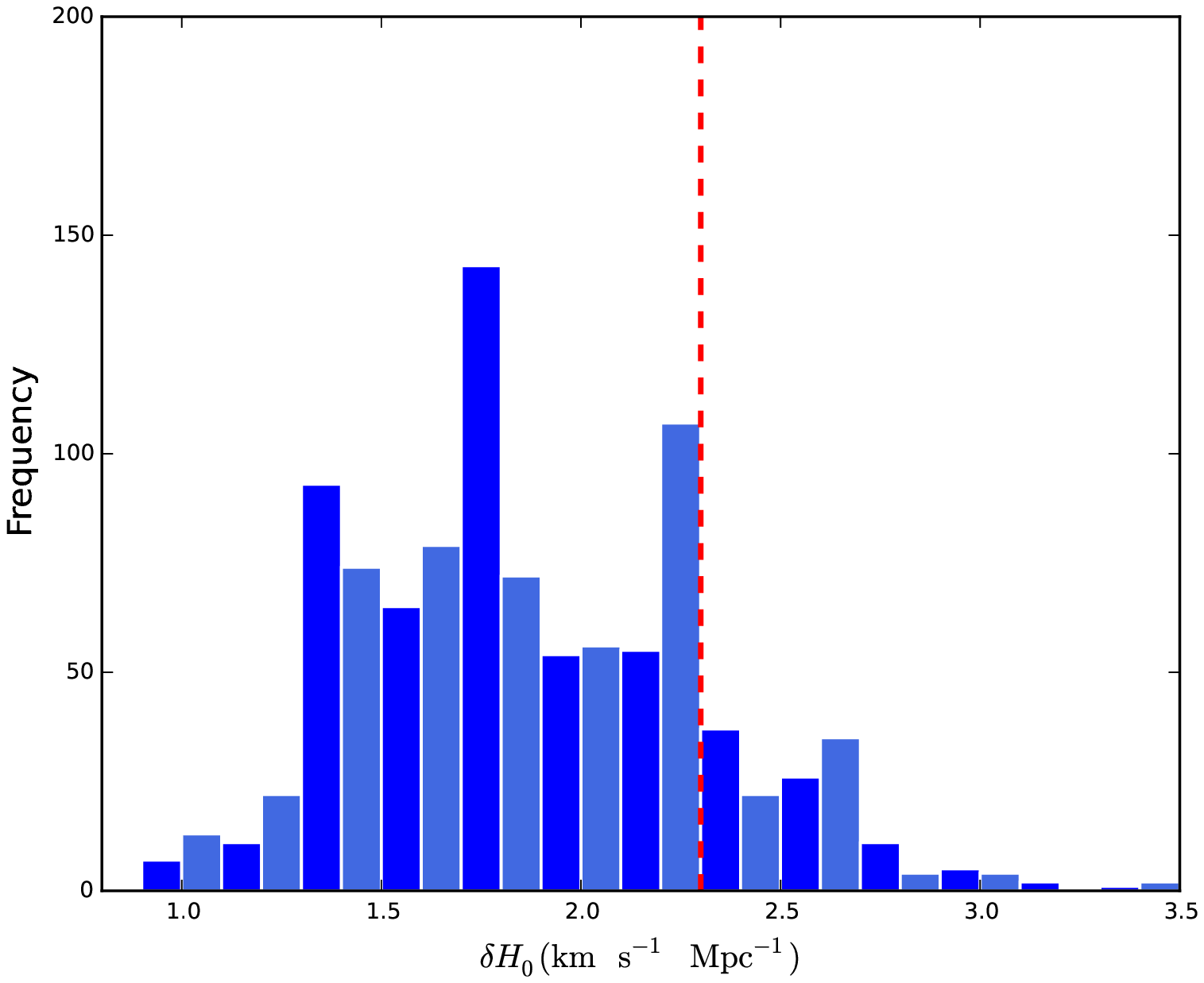}
\caption{ {\it Left panel}: The histogram of the $\delta H_0$ provided by 1000 {\it MC-gaussian-aniso} simulations. {\it Right panel}: the same histogram, but for the {\it MC-gaussian-iso} simulations. We find that the variance $\delta H_0 = 2.3 \; \mathrm{km \; s}^{-1} \mathrm{Mpc}^{-1}$, highlighted in the red dashed vertical line, or larger can be reproduced by $22.9$\% and $20.4$\% of these runs, respectively.} 
\label{fig:histograms_deltah_02}
\end{figure*}

The results of the statistical significance assessment for the $\delta H_0$ are exhibited in figures~\ref{fig:histograms_deltah_01} and~\ref{fig:histograms_deltah_02}. We find that it is possible to reproduce $\delta H_0 \geq 2.3 \; \mathrm{km \; s}^{-1} \mathrm{Mpc}^{-1}$ by $23.2$\% for the {\it MC-shuffle} and $14.7$\% for the {\it MC-iso} realisations, while the {\it MC-gaussian-aniso} and {\it MC-gaussian-iso} realisations give $22.9$\% and $20.4$\%, respectively. These results indicate that this variance presents moderate significance ($> 68.7$\% CL), and that it is most significant, albeit marginally, in those sets which assume isotropic distribution of data points. This result therefore suggests a possible bias in our analyses due to under and over-sampled patches in the sky with SNe events, as featured in figure~\ref{fig:SNe_maps}. The effect that a non-uniform SNe angular distribution could have on the cosmological parameters inference was previously studied by~\cite{hannestad}, who found that such anisotropies lead to larger uncertainties in their values, for instance. 

We also test whether the dipole direction of the simulated Hubble-maps depends on the SNe celestial distribution. For this purpose, the {\it MC-shuffle} and {\it MC-gauss-aniso} data sets are considered in this analysis as they assume the original distribution of data points in the sky. In practice, we compute how many of these realisations present a dipole lying in the $l \in [300^{\circ},330^{\circ}]$ and $b \in [12^{\circ},42^{\circ}]$ range of the celestial sphere, i.e., $\pm 15^{\circ}$ in both galactic coordinates with respect the actual dipole direction, besides the uncertainty related to the coarse pixelisation grid, which is $3.67^{\circ}$ in the case of $N_{side} = 16$. Such range would correspond to

\begin{equation}
\label{eq:p_acc_dip}
p_{acc} = \frac{\int_{l_1}^{l_2} dl \int_{b_1}^{b_2} db\sin{b}}{4\pi} \approx 1.36\% \;,
\end{equation} 

\noindent fraction of the whole sky, being $b_1$ and $b_2$ (same for the $l_1, l_2$) the lower and upper limit of this range, respectively~\footnote{This fraction in fact comprises roughly $2.0$\% of the discretised celestial sphere in the corresponding $N_{side}$}. Hence, if a dipole has equal probability of being detected in any patches in the sky, which must be true if there is no correlation between the axis of the maximal Hubble-maps asymmetry and the angular distribution of data points, the probability of finding such variance within this range should give about this value for both MC data sets. We obtain that about $2.5$\% of both MC prescriptions give a preferred maximal Hubble-map direction in this range, which thus support the hypothesis that the direction of the maximal $\delta H_0$ is not strongly affected by the sparsity presented in the SNe angular distribution. 

\subsection{Bias in the $H_0$} 

In addition, we test whether there is any bias in the $H_0$ measurement following the discussion of~\cite{romano}, who showed that the presence of inhomogeneities in the nearby structures (which could be responsible for the bulk flow motion) may lead to a higher value of this parameter than expected for a perfect FLRW Universe and hence a possible misinterpretation of the $H_0$ measurement in the low redshift range. In practice, we estimate the probability of obtaining a higher $H_0$ global fit in the presence of a non-negligible bulk flow velocity than the cases where no $v_{bf}$ is assumed a priori, and test how this probability is changed when the distance modulus error is reduced. Assuming $v_{bf} = 300 \mathrm{km \; s}^{-1}$ and $H_0 = 69.8 \; \mathrm{km \; s}^{-1} \mathrm{Mpc}^{-1}$, which correspond to the parameters best fit that we obtained for the Union2.1 data, we find that about $55$\% of the realisations with non-null bulk flow gives $H_0 \geq 69.8 \; \mathrm{km \; s}^{-1} \mathrm{Mpc}^{-1}$, whereas the probability of satisfying this hypothesis is $37$\% for the case with null bulk flow velocity. This result shows that, within the current uncertainty of the SNe distance measurements, it is not possible to demonstrate that a bulk flow in fact biases the $H_0$ determination. Nevertheless, such tension increases when the $\sigma_\mu$ of the SNe are reduced by the half or the quarter of their original values, where $28$\% of the realisations with null bulk flow satisfy the criterion above and $60$\% of those assuming $v_{bf} = 300 \mathrm{km \; s}^{-1}$ in the former case, while in the latter case $13$\% and $71$\%  (thus translated into a $1.6\sigma$ tension) of the null and non-null fiducial $v_{bf}$, respectively, for the same hypothesis. We conclude that there is indeed marginal evidence that the presence of such bulk flow affects the estimation of the Hubble Constant, agreeing with the results from~\cite{romano}, albeit this bias could only be detected if the observational uncertainties are significantly reduced. Nevertheless, such precision may be achieved with the much larger and more precise SNe data sets expected in the following years with the advent of the Euclid and LSST surveys.

%%%%%%%%%%%%%%%%%%%%%%%%%%%%%%%%%%%%%%%%%%%%%%%%
\section{Conclusions}    \label{conc}
%%%%%%%%%%%%%%%%%%%%%%%%%%%%%%%%%%%%%%%%%%%%%%%%

In this work, we constrained the variance of the Hubble Constant with the latest low-z SNe compilation available in the literature, investigating whether it could arise due to the bulk flow motion in our local Universe, and whether it could explain the $H_0$ tension reported by different probes. We performed our analyses adopting a hemispherical comparison procedure, whose hemispheres centres are determined by the HEALPix pixelisation scheme. We obtained a minimal $H_0$ of $68.9 \pm 0.5 \; \mathrm{km \; s}^{-1} \mathrm{Mpc}^{-1}$ and a maximal of $71.2 \pm 0.7 \; \mathrm{km \; s}^{-1} \mathrm{Mpc}^{-1}$, implying a local $H_0$ variance of $\delta H_0 = (2.30 \pm 0.83) \; \mathrm{km \; s}^{-1} \mathrm{Mpc}^{-1}$ considering $1\sigma$ uncertainties. These extreme $H_0$ values are in good agreement with the local and distant measurements of the Hubble Constant from~\cite{planck15} and~\cite{riess2}, respectively, and the variance amplitude is consistent with the variance due to the inhomogeneities in the low-z cosmic web estimated in previous works. Furthermore, we found that the $H_0$ variation through the celestial sphere is maximal towards $(l,b) = (315.00^{\circ}, 27.28^{\circ})$, which is close to the direction of the bulk flow velocity reported in the literature, thus indicating that this effect could indeed explain the $H_0$ variance, as well as the tension we have just discussed. 

We also found that the actual $\delta H_0$ value is significant at the $68.7$\% CL, while the isotropic cases present a slightly higher significance than those assuming the actual SNe angular distribution. These results point out, therefore, that this variance can be partially ascribed to the limitations of the available data set. We also found that its preferred direction does not show any significant dependence with respect to SNe under and over-sampled patches in the sky. Besides, we obtained moderate evidence that the presence of a bulk flow velocity leads to a higher $H_0$ value than the Union2.1 fit, but if the distance modulus uncertainties are reduced to a factor of 4. This result stresses, once again, that the current status of the SNe data cannot yet allow a significant estimation of the bulk flow motion biasing the $H_0$ parameter. A more precise assessment of such effect is left for a future work.

For now, the results obtained in this paper only provide moderate evidence that the SNe data is able to address the issue of the $H_0$ tension and its possible relation with the bulk flow motion, except for the well-determined direction for the maximal $H_0$ variance. However, we point out once again that the significance of these results is still limited by the current status of the SNe data set, especially related to its celestial incompleteness. As discussed by~\cite{math, bengaly}, it is difficult to estimate a dipolar signal in the SNe data when the angular distribution of these objects are highly non-uniform, so that we expect a more robust inference the $H_0$ directional analysis (as well as for the velocity flows) when larger and more precise data sets, with about $10^3-10^4$ SNe in $z \leq 0.1$, shall be available. This may be possibly detected with future sky surveys such as Euclid and LSST, for instance. Furthermore, there are some robust indications of large inhomogeneities in the local structures, as pointed out by~\cite{keenan}, who showed that the luminosity function of galaxies indicates the existence of a large inhomogeneity at $z \simeq 0.08$, which is consistent with results by~\cite{whitbourn}, besides,~\cite{app3} reported a potential hemispherical asymmetry on the $K$-band luminosity function using a galaxy catalogue in $z \leq 0.1$. All these results stress the importance of further studies using galaxy surveys in this low redshift range in order to perform a better inference of our motion through the Universe, thus allowing a better estimation of its impact on the cosmological parameters estimation, such as the $H_0$ value. In addition, these improvements may also provide a better assessment of some fundamental hypotheses of the cosmological concordance model, such as the Cosmological Principle, since knowing the extent and amplitude of the local structures dipole may enable us to discriminate between a real anisotropic (or inhomogeneous) signal and a local effect bias, for instance. 

\acknowledgments

The author thanks J. S. Alcaniz, R. S. Gon\c calves, A. E. Romano, besides the anonymous referee, for the highly valuable discussions. The author also thanks CAPES for the financial support, and acknowledges the HEALPix (http://healpix.jpl.nasa.gov) software package developed by~\cite{gorski} that made the analyses carried out in this work possible.

% ################### References ###################

\label{lastpage}

\end{document}